\documentclass
[a4paper,twoside,10pt]{letter}
\usepackage{graphicx,saj,multicol,subeqnarray}

\def\papertype{Preliminary report}

\def\udc{...}
\begin{document}
\baselineskip=3.1truemm
\columnsep=.5truecm
\newenvironment{lefteqnarray}{\arraycolsep=0pt\begin{eqnarray}}
{\end{eqnarray}\protect\aftergroup\ignorespaces}
\newenvironment{lefteqnarray*}{\arraycolsep=0pt\begin{eqnarray*}}
{\end{eqnarray*}\protect\aftergroup\ignorespaces}
\newenvironment{leftsubeqnarray}{\arraycolsep=0pt\begin{subeqnarray}}
{\end{subeqnarray}\protect\aftergroup\ignorespaces}
%


\markboth{\eightrm OPTICAL OBSERVATIONS OF IC342 GALAXY} {\eightrm M. M. VU\v{C}ETI\'{C} et al.}

{\ }

\publ

\type

{\ }


\title{OPTICAL OBSERVATIONS OF THE NEARBY GALAXY IC342 WITH NARROW BAND [S$\mathbf{II}$] AND H$\alpha$ FILTERS.
$\mathbf{II}$ -- Detection of 16 Optically-Identified Supernova Remnant Candidates\footnotemark[1] } \footnotetext[1]{Based on data
collected with the 2 m RCC telescope at Rozhen National Astronomical
Observatory}


\authors{M. M. Vu\v{c}eti\'{c}$^{1}$, A. \'{C}iprijanovi\'{c}$^{1}$,  M. Z. Pavlovi\'{c}$^{1}$, T. G. Pannuti$^{2}$, N. Petrov$^{3}$}
\authors{\"{U}. D. G\"{o}ker$^{4}$ and E. N. Ercan$^{4}$}

\vskip3mm


\address{$^{1}$Department of Astronomy, Faculty of Mathematics,
University of Belgrade,\break Studentski trg 16, 11000 Belgrade,
Serbia}

\Email{mandjelic}{math.rs}

\address{$^{2}$Department of Earth and Space Sciences, Space Science Center, Morehead State University, 235 Martindale Drive,  Morehead, KY 40351, USA}

\address{$^{3}$National Astronomical Observatory Rozhen, Institute of Astronomy, Bulgarian Academy of Sciences, 72 Tsarigradsko Shosse Blvd, BG-1784 Sofia,
Bulgaria}

\address{$^{4}$Department of Physics, Bo\u{g}azi\c{c}i University, North Campus, KB Building Floor 3-4, 34342, Istanbul, Turkey}


\dates{ , 2015}{ , 2015}


\summary{We present the detection of 16 optical supernova remnant (SNR) candidates in the nearby spiral galaxy IC342. The candidates were detected by applying \hbox{[S\,{\sc ii}]}/H$\alpha$ ratio criterion on observations made with the 2 m RCC telescope at Rozhen National Astronomical Observatory in Bulgaria. In this paper, we report the coordinates, diameters, H$\alpha$ and \hbox{[S\,{\sc ii}]} fluxes for 16 SNRs detected in two fields of view in the IC342 galaxy. Also, we estimate that the contamination of total H$\alpha$ flux from SNRs in the observed portion of IC342 is 1.4\%. This would represent the fractional error when the star formation rate (SFR) for this galaxy is derived from the total galaxy's H$\alpha$ emission.}


\keywords{ISM: supernova remnants -- Galaxies: individual: IC342.}
\vskip3mm
\begin{multicols}{2}
{


\section{1. INTRODUCTION}

In this paper, we present the second part of our study on the search for emission nebulae in the IC342 galaxy. In the first paper (Vu\v{c}eti\'{c} et al. 2013, hereafter Paper I) we presented the detection of 203 \hbox{H\,{\sc ii}} regions in the observed portion of this spiral galaxy. That raised the number of known \hbox{H\,{\sc ii}} regions 2.5 times in this part of IC342. In this paper, we give the details of the detection of 16 supernova remnant (SNR) candidates, out of which all except one represent SNR candidates detected for the first time.

IC342 is an almost face on spiral galaxy of large angular extent. It is heavily obscured by the Galactic disk, and that is why it is often avoided for optical observations. Also, due to the large extinction, until 1989 there was a large uncertainty in its distance, which ranged from 1.5 to 8 Mpc (McCall 1989). In this paper, we adopt 3.3 Mpc (Saha et al. 2002) which is a Cepheids distance to IC342. In Table 1, we give basic data on this galaxy.

Previous studies of emission nebulae in IC342 started with the work of D'Odorico et al. (1980), who were the first to search for IC342 SNRs at optical wavelengths. Their paper reported the detection of 4 SNRs, but they observed only the central part of the galaxy. Also, their paper did not give any flux measurements of detected objects. Afterwards, Hodge and Kennicutt (1983) in their atlas of \hbox{H\,{\sc ii}} regions detected 666 \hbox{H\,{\sc ii}} regions across the entire extent of the IC342 galaxy, but only the positions of the sources were given by these authors. Recently, Herrmann et al. (2008) performed an imaging survey using narrow band \hbox{[O\,{\sc iii}]} and H$\alpha$ filters to identify planetary nebulae: 165 such sources were found in this galaxy. As already mentioned, in Paper I we  presented the detection of 203 \hbox{H\,{\sc ii}} regions in the southwestern part of this galaxy through H$\alpha$ and \hbox{[S\,{\sc ii}]} filters.

Here we make mention on one very interesting object in this galaxy - IC342 X-1. It is one of the most studied ultraluminous X-ray (ULX) sources, with the optical ``Tooth'' nebula surrounding it. The Tooth nebula is probably either an SNR that reflects the formation of the compact star in the ULX or it is an X-ray
ionized bubble driven by strong outflows from the ULX. There are numerous papers (e.g. Roberts et al. 2003, Abolmasov et al. 2007, Feng and Kaaret 2008, Mak et al. 2011, Cseh et al. 2012, Marlowe et al. 2014) in which different kinds of interpretation of observations (optical, spectroscopic, X-ray and radio) of this source were done. This variable source has an average X-ray luminosity of $10^{40}$ erg/s. Recently, Marlowe et al. (2014) observed IC342 X-1 simultaneously in X-ray and radio domains with \textit{Chandra} and the Jansky Very Large Array (JVLA), respectively. The \textit{Chandra} data revealed a spectrum that
is well modeled by a thermal accretion disc spectrum, while no significant compact core radio emission was observed within the region of the ULX. On the other hand, extended radio emission with an estimated total flux density of $\sim$ 2 mJy at 5 GHz (VLA, Cseh et al. 2012) was found around the position of IC342 X-1, and its size is consistent with the size of optical nebula (280 pc $\times$ 130 pc).

In this study, the detection of SNRs was done using the fact that the optical spectra of SNRs have elevated \hbox{[S\,{\sc ii}]} $\lambda$671.7 nm, $\lambda$673.1 nm to H$\alpha$ $\lambda$656.3 nm emission-line ratios, as compared to the spectra of normal {\hbox{H\,{\sc ii}}} regions. This emission ratio is used to differentiate between shock-heated SNRs (ratios $>$0.4, but often considerably higher) and photoionized nebulae ($<$0.4, but typically $<$0.2) (Matonick and Fesen 1997; Blair and Long 1997). So far, more than 1200 optical SNRs or SNR candidates have been detected across 25 galaxies (see more details on optical SNRs in nearby galaxies in Vu\v{c}eti\'{c} et al. 2015). The M83 galaxy, with more than 300 optical SNRs (Blair and Long 2004, Dopita et al. 2010, Blair et al. 2012, Blair et al. 2014), is the best sampled for optical SNRs among all galaxies. Leonidaki et al. (2013) contributed more than 400 SNRs to the total number of optically detected SNRs in six nearby galaxies - NGC2403, NGC3077, NGC4214, NGC4395, NGC4449 and NGC5204. Almost all of these detections were done using \hbox{[S\,{\sc ii}]}/H$\alpha$ ratio criterion.

In this paper, we also discuss contamination of the total H$\alpha$ flux of this galaxy by SNRs. As for the derivation of SFR from  H$\alpha$ emission, only radiation from {\hbox{H\,{\sc ii}}} regions is relevant (see e.g. Kennicutt 1983), all other H$\alpha$ emitters should be removed in order to obtain appropriate SFRs. Vu\v{c}eti\'{c} et al. (2015) have shown, on the sample of galaxies that have been surveyed for optical SNRs, how flux coming from the SNRs affects the SFRs derived from H$\alpha$ flux. Similarly, Andjeli\'{c} (2011) has shown how H$\alpha$ derived SFRs for the Holmberg IX galaxy can be significantly changed if nebular emission from the ultraluminous X-ray sources is removed. Optical observations through narrow band H$\alpha$ and \hbox{[S\,{\sc ii}]} filters allow us to differentiate SNRs from {\hbox{H\,{\sc ii}}} regions, and in that way we can improve SFRs by removing the SNR contamination from the galaxies total H$\alpha$ flux.

\section{2. OBSERVATIONS AND DATA REDUCTION}

The observations were carried out on November 27-28 2011, with the 2 m Ritchey-Chr\'{e}tien-Coud\'{e} (RCC) telescope at the National Astronomical Observatory (NAO) Rozhen, Bulgaria ($\varphi = 41^\circ 41' 35'' ,\ \lambda = 24^\circ  44' 30'' ,\ h = 1759$ m). The telescope was equipped with VersArray: 1300B CCD camera with 1340$\times$1300 px array, with plate scale of 0\uu 257732/px (pixel size is 20 $\mu$m), giving the field of view $5'45''\times 5'35''$.\\

\begin{minipage}{\textwidth}
 {\bf Table 1.} Data for IC342 taken from NED$^{1}$.
\vskip3mm
\begin{tabular}{@{\extracolsep{0.0mm}}l r @{}}
\hline
 Right ascension (J2000) &  03h46m48s.5 \\
 Declination (J2000) & +68$^{\circ}$05$'$47$^{\prime\prime}$ \\
 Redshift & 0.000103 \\
 Velocity & 31 km s$^{-1}$\\
 Distance$^{2}$ & 3.3 Mpc \\
 Angular size &  $21.4' \times 20.9'$ \\
 Magnitude & 9.1 mag (B)\\
 Gal. extinction$^{3}$  & 2.024 mag (B)\\
\hline
\hline
\end{tabular}
 \vskip1mm
$^{1}${\footnotesize \texttt{http://ned.ipac.caltech.edu/}}\\
$^{2}${\footnotesize Saha et al. (2002)}\\
$^{3}${\footnotesize Schlafly and  Finkbeiner (2011)} \
\vskip 3mm
\end{minipage}
\vskip 3mm

We observed three fields of view (FOV) in IC342 (see Fig. 1 in Paper I). Centers of the two FOV, for which conditions were photometric, are: FOV1 --  R.A.(J2000) = 03:45:45.7, Decl.(J2000) = +68:04:11; FOV2 -- R.A.(J2000) = 03:46:49.9, Decl.(J2000) = +68:00:47.

The observations were performed with the narrow band \hbox{[S\,{\sc ii}]}, H$\alpha$ and red continuum filters. We took sets of three images through each
filter, with total exposure time of 2700s {for} each filter. Typical seeing was 1\uu 5 -- 2\uu 75. Standard star images, bias frames and sky flat-fields were also taken. Data reduction was done using standard procedures in IRAF\footnote{IRAF is distributed by the National Optical Astronomy Observatory, which is operated by the Association of Universities for Research in Astronomy, Inc., under cooperative agreement with the National Science Foundation.} and IRIS\footnote{Available from {\footnotesize \texttt{http://www.astrosurf.com/buil/}}}.

Images through each filter were firstly combined using the sigma-clipping method, then sky-subtracted and flux calibrated using the observations of the standard star Feige 34 from Massey et al. (1988). An astrometric reduction of the images was performed by using U.S. Naval Observatory's USNO-A2.0 astrometric catalogue (Monet et al. 1998). Afterwards, the continuum contribution was removed from the H$\alpha$ and \hbox{[S\,{\sc ii}]} images, and images were corrected for filter transmission. To obtain images which are absolutely flux-calibrated and contain only line emission, we also had to correct H$\alpha$ emission for
the contamination of \hbox{[N\,{\sc ii}]} lines at $\lambda$654.8 nm, $\lambda$658.3 nm (see Paper I for all details on the procedures of data reduction and flux calibration). For the detection of SNRs, we adopted the integrated (sum of both components) \hbox{[N\,{\sc ii}]}$\lambda 6548, 6583$/H$\alpha$ ratio of 1.07, as was measured by Roberts et al. (2003) in the Tooth nebula.

}
\end{multicols}



\centerline{\includegraphics[bb=0 0 600 516, keepaspectratio,width=16cm]{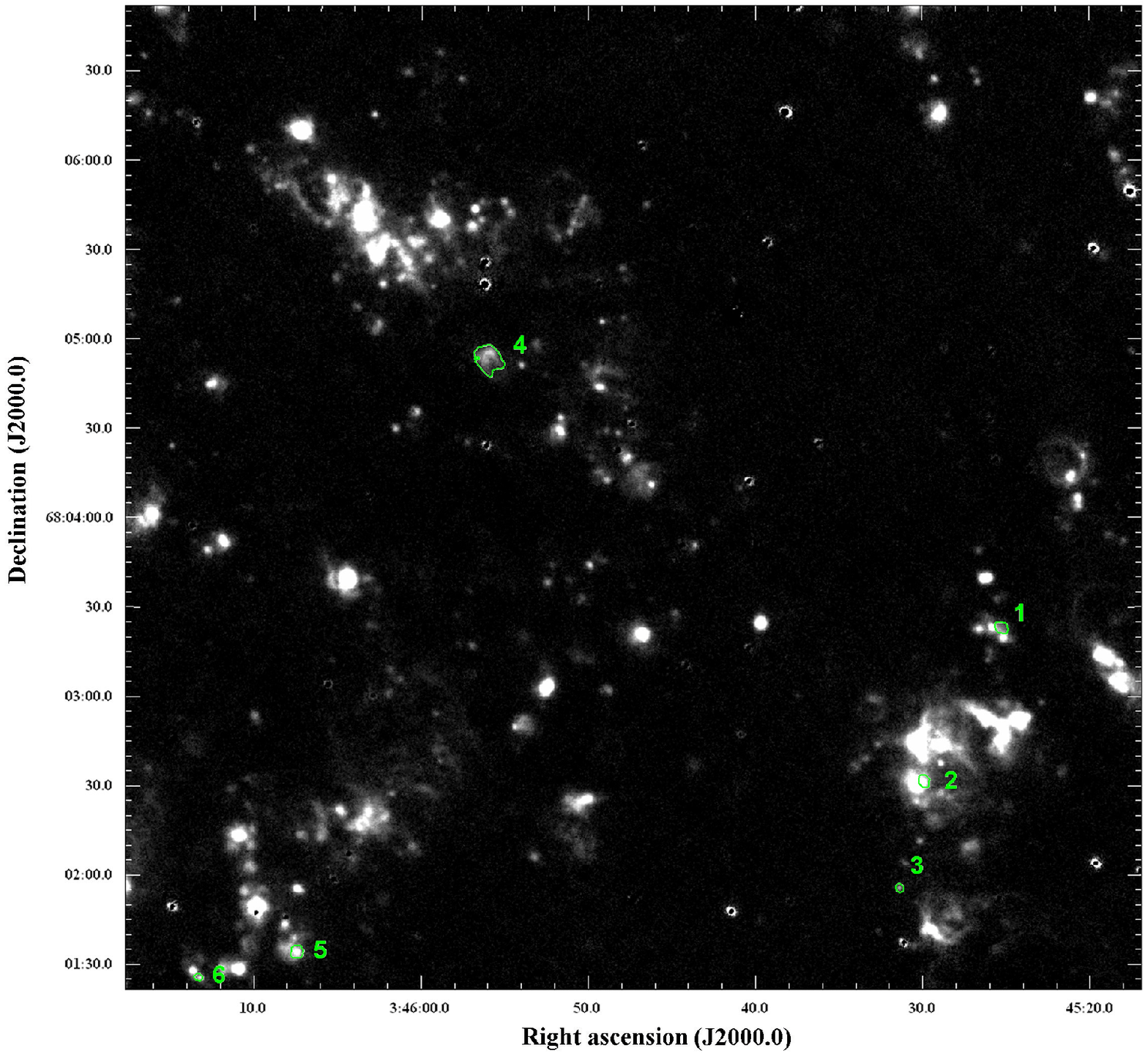}}


\vspace{1mm}
\begin{minipage}{14.5cm}
 \figurecaption{1.}{The continuum-subtracted H$\alpha$ image for FOV1 is identified for six sources. Numbers correspond to the entries in Table 2.}
\end{minipage}

\centerline{\includegraphics[bb=0 0 53 115, keepaspectratio,width=0.6\textwidth]{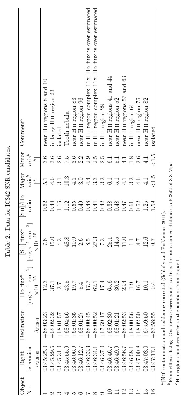}}

\centerline{\includegraphics[bb=0 0 600 516, keepaspectratio,width=16cm]{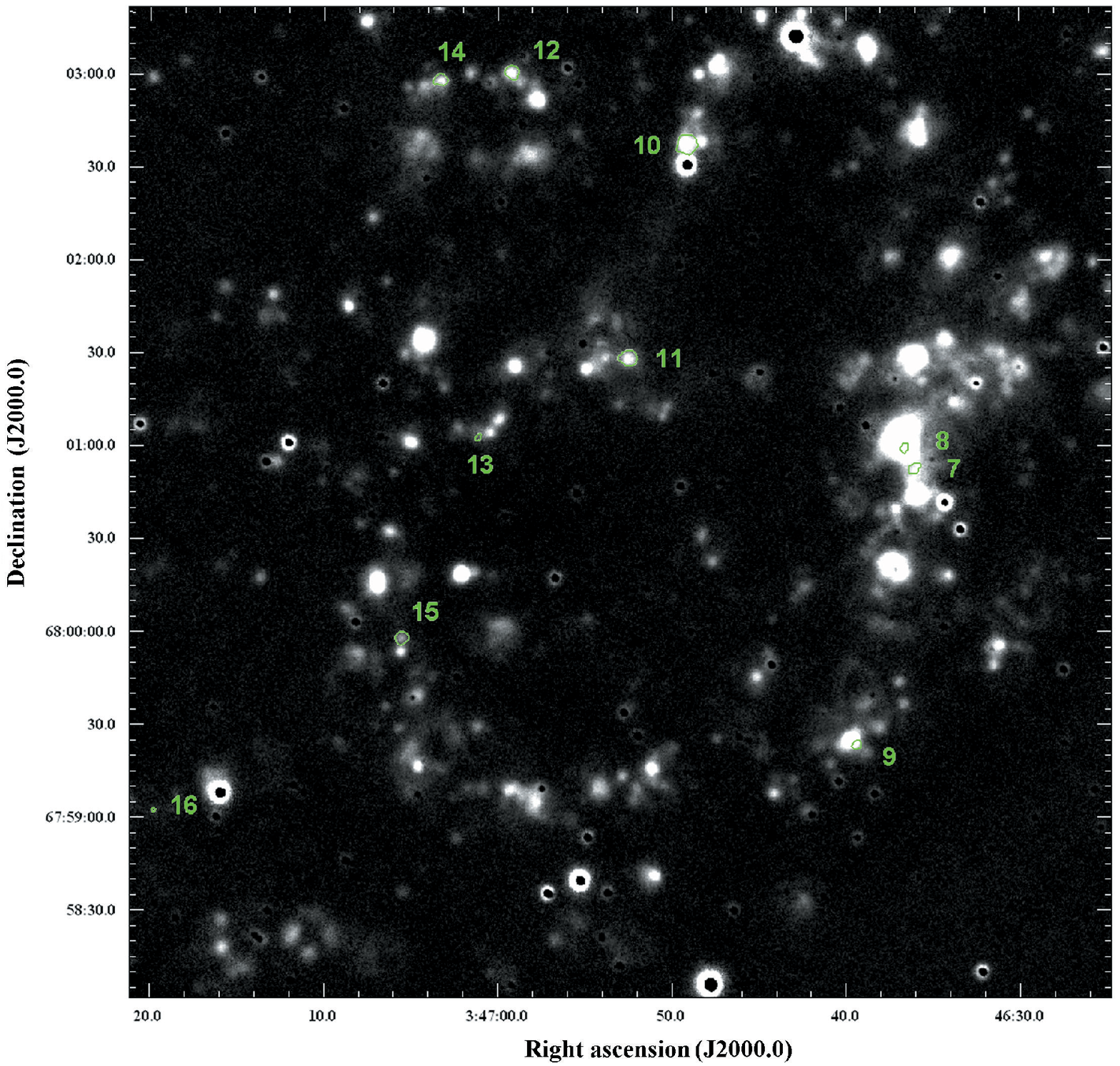}}


\vspace{1mm}
\begin{minipage}{14.5cm}
 \figurecaption{2.}{The continuum-subtracted H$\alpha$ image for FOV2 is identified for ten sources. Numbers correspond to the entries in Table 2.}
\end{minipage}

\begin{multicols}{2}

\section{3. RESULTS AND DISCUSSION}

In Table 2 we give the coordinates, diameters, H$\alpha$ and \hbox{[S\,{\sc ii}]} fluxes, and \hbox{[S\,{\sc ii}]}/H$\alpha$ ratios for 16 SNR candidates detected in two fields of view, observed in IC342. Only one object - the Tooth nebula, our object 4, is previously known, while the other 15 are new optical SNR candidates. As can be seen from Table 2, only three objects are isolated, while the remaining 13 are near or inside \hbox{H\,{\sc ii}} regions. This shows that our deep exposure observations are sensitive enough to detect SNRs in regions of higher density and to resolve possible confusion of SNRs with large HII regions. For those objects inside \hbox{H\,{\sc ii}} regions (objects number 2, 7, 8, 13), it is possible that fluxes given in Table 2 are over-estimated. In such cases it is difficult to differentiate SNR emission from emission coming from \hbox{H\,{\sc ii}} region.

In Paper I, we gave a detailed description of the procedure for extracting sources from the flux-calibrated image. First, we made an \hbox{[S\,{\sc ii}]} - 0.4H$\alpha$ image. All objects which were above zero level in this image and which did not have any emission in the continuum image, were considered as SNR candidates. For each member of this group, we placed a contour around the object on the \hbox{[S\,{\sc ii}]} image in order to extract the emission which is 2.5$\sigma$ above the background level. Then we used the same contours for each object to measure H$\alpha$ flux from the H$\alpha$ image. For the four previously mentioned objects located inside \hbox{H\,{\sc ii}} regions, we extracted emission only from the region which is bright in the \hbox{[S\,{\sc ii}]} - 0.4H$\alpha$ image. In Figs. 1 and 2, H$\alpha$ images with marked SNR candidates are shown. Sources are marked with contours which extract emission 2.5$\sigma$ above the background level on \hbox{[S\,{\sc ii}]} image. Positions and diameters of the objects were measured by fitting an ellipse to the outer source contour, using the SAOImage DS9 package. Typical errors of the positions of the sources is about one arc second.

In order to check our photometry calibration, we compared our estimate of H$\alpha$ flux of the Tooth nebula (our object 4) with previously published values. We found that our reddening-corrected H$\alpha$ flux, which is 4.3 $\times 10^{-14}$ erg s$^{-1}$ cm$^{-2}$, is between values published by Abolmasov et al. (2007) and Feng and Kaaret (2008). H$\alpha$ flux of Tooth nebula published by Roberts et al. (2003) is ten times lower than our flux, but as suggested by Feng and Kaaret (2008), we believe that flux from Roberts et al. (2003) which is quoted as reddening-corrected is not corrected for extinction. Also, we calculated \hbox{[S\,{\sc ii}]}/H$\alpha$ ratio in Tooth nebula to be 1.12, which is consistent with values 1.09 and 1.07, found in Roberts et al. (2003) and Abolmasov et al. (2007), respectively.

As already mentioned, D'Odorico et al. (1980) published the detection of four optical SNRs in the central part of this galaxy. Our observations covered only parts of the galaxy where objects SNR2 and SNR3 from D'Odorico et al. (1980) are located. Our analysis disclaimed both of these objects as SNR candidates, because none of them turned bright on our \hbox{[S\,{\sc ii}]} - 0.4H$\alpha$ image. After visual inspection of Figure 5 from D'Odorico et al. (1980), and comparison with our images, we believe that SNR2 is most probably \hbox{H\,{\sc ii}} region 125, while SNR3 could be \hbox{H\,{\sc ii}} region 138 from Paper I.

Herrmann et al. (2008) detected 165 PNe in IC342, but they used only photometric data of detected objects, so we could not check whether any of our SNR candidates match any PNe.

We also performed cross-correlation of our objects, with available X-ray and radio data for this galaxy. The only available radio observation of the whole IC342 galaxy was published by Baker et al. (1977), but none of the objects they detected matched any of our SNR candidates. On the contrary, this galaxy has been observed frequently in X-rays -- by \emph{Einstein}, \emph{ROSAT}, \emph{Chandra} and \emph{XMM-Newton} telescopes (Fabbiano and Trinchieri 1987, Bregman et al. 1993, Mak et al. 2008, Kong 2003, Bauer et al. 2003, Evans et al. 2010, Liu 2011, Mak et al. 2011). A systematic near-position search, with a search radius of 6$^{\prime\prime}$, revealed IC342 X-1 (Tooth nebula) as the only object from our Table 2 which was also detected in X-rays. Our object 4 is separated 1\uu7 from IC342 X-1, as it is catalogued in The \textit{Chandra} Source Catalog (Evans et al. 2010).

From Table 2 we find that the sum of H$\alpha$ fluxes for our 16 SNR candidates is 3.67 $\times 10^{-13}$ erg s$^{-1}$ cm$^{-2}$. From our
continuum-subtracted H$\alpha$ images we measured total H$\alpha$ flux of both FOV1 and FOV2 to be 26.1 $\times 10^{-12}$ erg s$^{-1}$ cm$^{-2}$. This gives us that H$\alpha$ flux from SNRs represents 1.4\% of total H$\alpha$ flux in the observed part of the galaxy. If we extrapolate this ratio of H$\alpha$ flux coming from SNRs to total H$\alpha$ flux for the whole IC342 galaxy, following Vu\v{c}eti\'{c} et al. (2015) we can say that SNR contamination of derived SFR for this galaxy is 1.4\%. This percentage is only a lower limit on SNR contamination of derived SFR, because of numerous observational selections effects which are present in optical detection of SNRs. On the other hand, the assumption that this SNR to total H$\alpha$ ratio is spread over the whole IC342 galaxy is rather rough and incorrect. It is taken here only for the purpose of a first estimate.

\section{4. CONCLUSIONS}

In this paper we present the properties of 16 SNR candidates in the nearby spiral galaxy IC342. Of these 16 potential SNRs, classified on their \hbox{[S\,{\sc ii}]}/H$\alpha$ ratios, 15 have been detected for the first time in this work. We show that objects designated as SNR2 and SNR3 in D'Odorico et al. (1980) are most likely not SNRs.

The contribution of the H$\alpha$ flux from the SNRs to the total H$\alpha$ flux and its influence on the estimate of SFR for IC342 are discussed. We find that SNRs contribute 1.4$\%$ of the contaminating H$\alpha$ flux in the observed portion of IC 342: this cause the error if SFR for this galaxy would be derived from the total galaxy's H$\alpha$ emission.

Our future observations will cover the full extent of IC342 galaxy and reveal final status of emission nebulae in this galaxy. This will also give us the opportunity to make a good assessment of the contamination of the total H$\alpha$ flux of this galaxy by SNRs, allowing for correction to the derived SFR in IC342.


\acknowledgements{This research has been supported by the Ministry of Education, Science {and Technological Development} of the Republic of Serbia through the project No. 176005 ``Emission nebulae: structure and evolution'' and it is a part of the joint project of Serbian Academy of Sciences and Arts and Bulgarian Academy of Sciences ``Optical search for supernova remnants and {\hbox{H\,{\sc ii}}} regions in nearby galaxies (M81 group and IC342)''. Authors gratefully acknowledge observing grant support from the Institute of Astronomy and Rozhen National Astronomical Observatory, Bulgarian Academy of Sciences. The authors \"{U}. D. G\"{o}ker and E. N. Ercan would like to thank for their financial support to Bo\v{g}azi\c{c}i University through the BAP Scientific Research Project No. 8563 ``The Theoretical Modelling of the Formation and Evolution of Supernova Remnants and Supporting with the Observations'' and T\"{U}BITAK (The Scientific and Technological Research Council of Turkey) under project code 113F117.}


\references

Abolmasov, P., Fabrika, S., Sholukhova, O. and Afanasiev, V.: 2007, \journal{Astrophysical Bulletin}, \vol{62}, 36.

Andjeli\'{c}, M. M.:2011, \journal{Serb. Astron. J.}, \vol{183}, 71.

Baker, J. R., Haslam, C.G.T., Jones, B. B. and Wielebinski, R.: 1977, \journal{Astron. Astrophys.}, \vol{59}, 261.

Bauer, F. E., Brandt, W. N. and  Lehmer, B.: 2003, \journal{Astron. J.}, \vol{126}, 2797.

Blair, W. P. and Long, K. S.: 1997, \journal{Astron. Astrophys. Suppl. Ser.}, \vol{108}, 261.

Blair, W. P. and Long, K. S.: 2004, \journal{Astrophys. J. Suppl. Ser.}, \vol{155}, 101.

Blair, W. P., Winkler, P. F. and Long K. S.: 2012, \journal{Astrophys. J. Suppl. Ser.}, \vol{203}, 8.

Blair, W. P. et al.: 2014, \journal{Astrophys. J.}, \vol{788}, 55.

Bregman, J. N., Cox, C. V. and Tomisaka, K.: 1993, \journal{Astrophys. J.}, \vol{415}, L79.

Cseh, D. et al.: 2012, \journal{Astrophys. J.}, \vol{749}, 17.

D'Odorico, S., Dopita, M. A. and Benvenuti, P.: 1980, \journal{Astron. Astrophys. Suppl. Ser.}, \vol{40}, 67.

Dopita, M. A. et al.: 2010,  \journal{Astrophys. J.}, \vol{710}, 964.

Evans, I. N. et al.: 2010,  \journal{Astron. Astrophys. Suppl. Ser.}, \vol{189}, 37.

Fabbiano, G. and Trinchieri, G.: 1987, \journal{Astrophys. J.}, \vol{315}, 46.

Feng, H. and Kaaret, P.: 2008, \journal{Astrophys. J.}, \vol{675}, 1067.

Herrmann, K. A., Ciardullo, R., Feldmeier, J. J. and Vinciguerr, M.: 2008, \journal{Astrophys. J.}, \vol{683}, 630.

Hodge, P. W. and Kennicutt, R. C. Jr.: 1983, \journal{Astron. J.}, \vol{88}, 296.

Kennicutt, R. C. Jr.: 1983, \journal{Astrophys. J.}, \vol{272}, 54.

Kong, A.K.H.: 2003, \journal{Mon. Not. R. Astron. Soc.}, \vol{346}, 265.

Leonidaki, I., Boumis, P. and Zezas A.: 2013, \journal{Mon. Not. R. Astron. Soc.}, \vol{429}, 189.

Liu, J.: 2011, \journal{Astrophys. J. Suppl. Series}, \vol{192}, 10.

Mak, D. S. Y. et al.: 2008, \journal{Astrophys. J.}, \vol{686}, 995.

Mak, D. S. Y., Pun, C. S. J. and Kong, A. K. H.: 2011, \journal{Astrophys. J.}, \vol{728}, 10.

Marlowe, H. et al.: 2014, \journal{Mon. Not. R. Astron. Soc.}, \vol{444}, 642.

Massey, P., Strobel, K., Barnes, J. V., and Anderson, E.: 1988, \journal{Astrophys. J.}, \vol{328}, 315.

Matonick,  D. M. and Fesen, R. A.: 1997, \journal{Astrophys. J. Suppl. Series}, \vol{112}, 49.

McCall, M. L.: 1989, \journal{Astron. J.}, \vol{97}, 1341.

Monet, D. et al.: 1998, USNO-A2.0 - A catalog of astrometric standards, U.S. Naval Observatory ({\texttt{http://tdc-www.harvard.edu/catalogs/ ua2.html}}).

Roberts, T. P., Goad, M. R., Ward, M. J. and Warwick, R. S.: 2003, \journal{Mon. Not. R. Astron. Soc.}, \vol{342}, 709.

Saha, A., Claver, J. and Hoessel, J. G.: 2002, \journal{Astron. J.}, \vol{124}, 839.

Schlafly, E. F. and  Finkbeiner, D. P.: 2011, \journal{Astrophys. J.}, \vol{737}, 103.

Vu\v{c}eti\'{c}, M. M., Arbutina, B., Uro\v{s}evi\'{c}, D., Dobard\v{z}i\'{c}, A., Pavlovi\'{c}, M. Z., Pannuti, T. G. and Petrov, N.: 2013, \journal{Serb. Astron. J.}, \vol{187}, 11 (Paper I).

Vu\v{c}eti\'{c}, M. M., Arbutina, B., Uro\v{s}evi\'{c}, D.: 2015, \journal{Mon. Not. R. Astron. Soc.}, \vol{446}, 943.

\endreferences

\end{multicols}

\vfill\eject

{\ }



\naslov{\noindent OPTIQKA POSMATRANJA BLISKE GALAKSIJE $\mathbf{IC342}$
KROZ USKE $\mathbf{[SII]}$ I $\mathbf{H}\alpha$ FILTERE.
$\mathbf{II}$ -- DETEKCIJA 16 OPTIQKI IDENTIFIKOVANIH KANDIDATA ZA OSTATKE SUPERNOVIH}



\authors{M. M. Vu\v{c}eti\'{c}$^{1}$, A. \'{C}iprijanovi\'{c}$^{1}$,  M. Z. Pavlovi\'{c}$^{1}$, T. G. Pannuti$^{2}$, N. Petrov$^{3}$}
\authors{\"{U}. D. G\"{o}ker$^{4}$ and E. N. Ercan$^{4}$}

\vskip3mm


\address{$^{1}$Department of Astronomy, Faculty of Mathematics, University of Belgrade,\break Studentski trg 16, 11000 Belgrade, Serbia}

\Email{mandjelic}{math.rs}

\address{$^{2}$Department of Earth and Space Sciences, Space Science Center, Morehead State University, Morehead, KY 40351, USA}

\address{$^{3}$National Astronomical Observatory Rozhen, Institute of Astronomy, Bulgarian Academy of Sciences, 72 Tsarigradsko Shosse Blvd, BG-1784 Sofia, Bulgaria}

\address{$^{4}$Department of Physics, Bo\u{g}azi\c{c}i University, North Campus, KB Building Floor 3-4, 34342, Istanbul, Turkey}

\vskip.7cm


\centerline{UDK \udc}


\centerline{\rit Prethodno saop\ss tenje}

\vskip.7cm

\begin{multicols}{2}
{


{\rrm U radu je prezentovana detekcija 16 optiqkih ostataka supernovih (OSN) u obli\zz njoj spiralnoj
galaksiji $\mathrm{IC342}$. Detekcija je izvr\ss ena upotrebom kriterijuma vezanog za odnos $\mathrm{[SII]}$ i
$\mathrm{H}\alpha$ linija, koriste\cc i posmatranja sa dvometarskog teleskopa Nacionalne astronomske opservatorije Ro\zz
en u Bugarskoj. U dva posmatrana vidna polja detektovano je ukupno 16 OSN
qiji su polo\zz aji, dijametri, kao i $\mathrm{H}\alpha$ i $\mathrm{[SII]}$ fluksevi navedeni u radu. Takodje, procenjeno je da je kontaminacija $\mathrm{H}\alpha$ fluksa fluksom koji potice sa OSN u posmatranom delu galaksije 1.4\%. Ovaj procenat bi predstavljao gre\ss ku kada bi se stopa formiranja zvezda u ovoj galaksiji odredjivala iz ukupnog $\mathrm{H}\alpha$ fluksa galaksije. }

}
\end{multicols}

\end{document}